\RequirePackage{fix-cm}
\documentclass[twocolumn,epjc3]{svjour3}  
\smartqed  
\RequirePackage{graphicx}
\RequirePackage{bm}
\RequirePackage{mathrsfs}
\RequirePackage{amsmath}
\RequirePackage{cancel}
\RequirePackage{amssymb}
\RequirePackage{url}
\journalname{Eur. Phys. J. C}
\RequirePackage{amsmath}
\RequirePackage{cases}
\begin{document}

\title{Comments on Gravitoelectromagnetism of Ummarino and Gallerati in ``Superconductor in a weak static gravitational field'' vs Other Versions}



\author{Harihar Behera\thanksref{e1,addr1}}
%

\thankstext{e1}{e-mail: behera.hh@gmail.com}


\institute{Physics Department, BIET Higher Secondary School, Govindpur, Dhenkanal-759001, Odisha, India \label{addr1}}

\date{Received: date / Accepted: date}
\maketitle
\begin{abstract}
Recently reported [Eur. Phys. J. C., {\bf 77}, 549 (2017)] gravitoelectromagnetic equations of Ummarino and Gallerati (UG) in their linearized version of General Relativity (GR) are shown to match with (a) our previously reported special relativistic Maxwellian Gravity equations in the non-relativistic limit and with (b) the non-relativistic equations derived here, when the speed of gravity $c_g$ (an undetermined parameter of the theory here) is set equal to $c$ (the speed of light in vacuum). Seen in the light of our new results, the UG equations satisfy the Correspondence Principle (cp), while many other versions of linearized GR equations that are being (or may be) used to interpret the experimental data defy the cp. Such new findings assume significance and relevance in the contexts of recent detection of gravitational waves and the gravitomagnetic field of the spinning earth and their interpretations. Being well-founded and self-consistent, the equations may be of interest and useful to researchers exploring the phenomenology of gravitomagnetism, gravitational waves and the novel interplay of gravity with different states of matter in flat space-time like UG's interesting work on superconductors in weak gravitational fields.
\end{abstract}
\section{Introduction}
\label{intro}
In a recent interesting theoretical study on the interplay of superconductivity and weak static gravitational field, Ummarino and Gallerati \cite{1} concluded that the reduction of the gravitational field in a superconductor, if it
exists, is a transient phenomenon and depends strongly on the parameters that characterize the superconductor. The gravitational equations used by the authors in their study are represented by the following Gravito-Maxwell Equations:
\begin{eqnarray}
\mathbf{\nabla}\cdot \mathbf{E}_g = -\, 4\pi G \rho_g,  \\
\mathbf{\nabla}\cdot \mathbf{B}_g = 0,       \\ 
\mathbf{\nabla}\times \mathbf{B}_g = -\,\frac{4\pi G}{c_g^2}\mathbf{j}_g\,+\,\frac{1}{c_g^2}\frac{\partial\mathbf{E}_g}{\partial t}, \\
\mathbf{\nabla}\times \mathbf{E}_g = -\,\frac{\partial\mathbf{B}_g}{\partial t},
\end{eqnarray}
\noindent
and the equation of motion of a particle moving with non-relativistic velocity, $\mathbf{v}$, is given by Gravito-Lorentz force law: 
\begin{equation}
\frac{d\mathbf{v}}{dt}\,=\,\mathbf{E}_g \,+\,\mathbf{v}\times\mathbf{B}_g.
\end{equation}
In the above equations $\rho_g = \rho_0$ is the (rest) mass density, $\mathbf{j}_g = \rho_g \mathbf{v}$ mass current density, the speed of gravitational waves in vacuum $c_g = c $ (the speed of light in vacuum) in \cite{1}, $\mathbf{E}_g$ is the usual Newtonian gravitational field (called gravitoelectric field) and $\mathbf{B}_g$ is the gravitational analogue of magnetic induction field (called gravito-magnetic field). The $\mathbf{E}_g$ and $\mathbf{B}_g$ fields are related to gravitoelectric scalar potential $\phi_g$ and gravitomagnetic vector potential $\mathbf{A}_g$ as 
\begin{eqnarray}
\mathbf{E}_g\,=\,-\,\mathbf{\nabla}\phi_g\,-\,\frac{\partial \mathbf{A}_g}{\partial t}, \\
\mathbf{B}_g\,=\,\mathbf{\nabla}\times \mathbf{A}_g.
\end{eqnarray}  
Ummarino and Gallerati \cite{1} derived these equations from Einstein's GR by linearization procedure in the weak field and slow motion approximations. We name the equations (1-7) to represent the General Relativistic Maxwellian Gravity of Ummarino-Gallerati as (GRMG-UG) to differentiate it from other existing (or future) formulations that (may) result from different methods of study. For instance, in this communication we show how one can derive these equations in a non-relativistic approach without invoking the space-time curvature and the linearization schemes of GR. By adopting, in essence, the non-relativistic approach of Schwinger et al. \cite{2} for their derivation of Maxwell's equations and the Lorentz force law in the electromagnetic case, we derive the fundamental equations of Time-Dependent Galileo-Newtonian Gravitodynamics of moving bodies (called here as Non-Relativistic Maxwellian Gravity or NRMG in short) within the Galileo-Newtonian relativity physics by combining the following ingredients: \\
(a) The validity of Newton's laws of gravitostatics, \\
(b) The validity of the equation of continuity that expresses the law of conservation of mass\footnote{Schwinger et al. \cite{2} derived (b) using the Galileo-Newton principle of relativity (masses at rest and masses with a common velocity viewed by a co-moving observer are physically indistinguishable). Here, we will use this relativity principle for deriving the gravitational analogue of the Lorentz force law.},  and \\
(c) Postulating the existence of gravitational waves traveling in free space with a finite velocity $c_g$ (which is an undetermined parameter of the theory but whose value may be fixed in measurement of physical quantities involving $c_g$ or by comparing the field equations with those obtainable from more advanced theories). \\
\indent
As will be shown here, our derived equations of NRMG match with the equations (1-7) of GRMG-UG, if $c_g = c$. Further, if $c_g = c$, the equations of NRMG also match with those of Special Relativistic Maxwellian Gravity (SRMG) \cite{3} in flat space-time, where $c_g = c$ is a natural outcome of the theory there (here discussed in Sec. 3). From McDonald's \cite{4} report of little known Heaviside's Gravity (HG) \cite{4,5} of 1893, we find that the equations of NRMG also match with those of HG in which Heaviside thought $c_g$ might be equal to $c$. These findings assume significance and relevance in the contexts of recent experimental detection of gravitational waves \cite{6,7,8,9}, gravito-magnetic field of the spinning Earth using the orbital data of two laser-ranged satellites (LAGEOS and LAGEOS II) \cite{10,11,12,13,14} and the Gavity Probe B (GP-B) experimental results \cite{15,16,17} vindicating Einstein's GR, because all of these results are being interpreted in the literature as new crucial tests of general relativity having no Galileo-Newtonian counterpart. For instance, Ciufolini et al. \cite{10}, in their report of measurement of the Lense-Thirring effect\footnote{This effect and the precessing spin axis of on board gyroscopes of the GP-B experiment etc. can also be understood in terms of gravitomagnetic effects.} stated:\\
{\it Newton's law of gravitation has a formal counterpart in Coulomb's law of
electrostatics; however, Newton's theory has no phenomenon formally analogous to
magnetism. On the other hand, Einstein's theory of gravitation predicts that the force
generated by a current of electrical charge, described by Amp\`{e}re's law, should also have
a formal counterpart ``force" generated by a current of mass.}
\section{Non-Relativistic Maxwellian Gravity (NRMG)}
\label{sec:2}
In Galileo-Newtonian physics, gravitational mass\footnote{The mass that appears in Newton's law of Gravitation is called the gravitational mass of a body.}, $m_g$, is the source of Newtonian gravitational field, $\mathbf{E}_g$, which obeys the following two equations, viz., 
\begin{equation}
\mathbf{\nabla}\cdot\mathbf{E}_g\,=\,-\,4\pi G\rho_g \qquad {\mbox{and}}
\end{equation}
\begin{equation}
\mathbf{\nabla}\times \mathbf{E}_g\,=\,\vec{0}, 
\end{equation}
\noindent
where $\rho_g =\rho_0$ is the (positive) rest mass density (to be shown shortly in this section and also in the next section in the relativistic case) and $G$ is Newton's universal gravitational constant having the value  $6.674\times 10^{-11}$ N$\cdot$m$^2$/kg$^2$. In an inertial frame, the gravitational force on a point particle having gravitational mass $m_g$ in a gravitational field $\mathbf{E}_g$ is expressed as 
\begin{equation}
\mathbf{F}_g = m_g\mathbf{E}_g = -m_g \mathbf{\nabla}\phi_g,  
\end{equation}
\noindent
where $\phi_g$ is gravitational potential at the position of $m_g$. Further, the mass that appears in Newton's second law of motion, which Newton actually wrote in The Principia, is of the following form (valid only in inertial frames):
\begin{equation}
\frac{d}{dt}(m\mathbf{v})\,=\,\vec{F}={\mbox{Net force on}}\, m, 
\end{equation}
is called the inertial mass, $m$, which is a measure of inertia of a body's resistance to any change in its state of motion represented by its velocity $\mathbf{v}$ or linear momentum $\vec{p} = m\vec{v}$. If $m$ does not change with time, then the most powerful and profound equation (11) of Newton in classical  physics, whose form survived special relativistic revolution, takes the familiar non-relativistic ($v << c$) form,
\begin{equation}
m\frac{d\vec{v}}{dt}\,=\,m\vec{a}\,=\,\vec{F}, 
\end{equation}
where $\vec{a}$ is the acceleration of $m$ as measured in an inertial frame. 
In Galileo-Newtonian physics, inertia is of two types, viz., inertia of rest and inertia of motion. Accordingly, inertial mass is of two types: rest mass ($m_0$ = a measure of inertia of a body at rest) and inertial\footnote{Despite the fact that rest mass is also inertial mass, we are reluctant to add one more adjective to mass in motion.} mass ($m$ = a measure of inertia of a body in motion). Thus a qualitative distinction between two types of inertial masses exists in Newtonian physics, but there is no quantitative distinction, i.e., $m = m_0$ in Newtonian world of physics\footnote{However, this is not the case in relativistic world of physics. But we are least concerned about this in the non-relativistic situation that we want to study in this section. In the next section the equality of $m_g = m_0$ will be clearly demonstrated in a thought experiment without violating Galileo's law of universality of free fall in the relativistic case also.}. With this understanding, if the only force acting on the particle is the gravitational force, then the equation of motion of a particle of rest mass $m_0$ and gravitational mass $m_g$ can be obtained from equations (10) and (12) as 
\begin{equation}
\frac{d\mathbf{v}}{dt}\,=\,\frac{m_g}{m_0}\mathbf{E}_g.
\end{equation}  
Equation (13) shows us that if $m_g = m_0$ (or $\rho_g = \rho_0$ for continuous mass distribution), then any particle of whatever rest mass will fall with the same acceleration in a given gravitational field. This is the law of universality of free fall - known to be true since the time of Galileo. 
In view of these observations, in this section term `mass' is to be understood as the rest mass which represents the gravitational charge (or mass) of a particle. \\ 
In Newtonian physics, mass obeys a local conservation law expressed by the equation of continuity:
\begin{equation}
\mathbf{\nabla}\cdot \mathbf{j}_0(\mathbf{r}, t)\,+\,\frac{\partial }{\partial t}\rho_0(\mathbf{r},t)\,=\,0.
\end{equation} 
\noindent
where $\mathbf{j}_0 = \rho_0(\mathbf{r},t) \mathbf{v}$ is the rest mass current density. With $\rho_g = \rho_0$, the three laws expressed in equations (8), (9) and (14) have their individual indisputable validity in Newtonian Physics. But they are simultaneously valid only for the systems or situations where 
\begin{subequations}
\begin{align}
\mathbf{\nabla}\cdot \mathbf{j}_0(\mathbf{r}, t)\,=\,0, \\
\frac{\partial }{\partial t}\rho_0(\mathbf{r},t)\,=\,0, \\
\frac{\partial }{\partial t}\mathbf{E}_g(\mathbf{r}, t)\, =\,\vec{0},
\end{align}
\end{subequations}
remain valid (here in (15) the time dependence of the physical quantities is made explicit). The equations (8-15) describe the familiar Galileo-Newtonian gravitodynamics. Newton's gravitational force law (10) implies ``action-at-a-distance''- the gravitational force seems to act instantaneously at a distance - which Newton considered as an absurd element of his theory. Despite his great efforts, Newton could not offer a plausible mechanism for resolving the problem of action-at-a-distance and left the problem for others to resolve as evident from his 3rd letter to Bentley, dated February $25, 1692/3$, where he wrote \cite{18}:
\begin{quote}
It is inconceivable, that inanimate brute Matter should, without the Mediation of something else, which is not material, operate upon, and affect other Matter without mutual Contact, as it must be, if Gravitation in the Sense of \emph{Epicurus}, be essential and inherent in it. And this is one Reason why I desired you would not ascribe innate Gravity to me. That Gravity should be innate, inherent and essential to Matter, so that one Body may act upon another at a Distance thro' a \emph{Vacuum}, without the Mediation of any thing else, by and through which their Action and Force may be conveyed from one to another, is to me so great an Absurdity, that I believe no Man who has in philosophical Matters a competent Faculty of thinking, can ever fall into it. Gravity must be caused by an Agent acting constantly according to certain Laws; but whether this Agent be material or immaterial, I have left to the Consideration of my Readers.
\end{quote}
In our attempt to resolve Newton's action-at-a-distance problem within his domain of physics, we first wish to explore a system where the Gauss's law of gravitostatics (8) and the equation of continuity (14) work peacefully or co-exist simultaneously but now with the restrictions in (15) removed by the following conditions:
\begin{subequations}
\begin{align}
\mathbf{\nabla}\cdot \mathbf{j}_0(\mathbf{r}, t)\,\neq \,0, \\
\frac{\partial }{\partial t}\rho_0(\mathbf{r},t)\,\neq \,0, \\
\frac{\partial }{\partial t}\mathbf{E}_g(\mathbf{r}, t)\, \neq \, \vec{0}.
\end{align}
\end{subequations}
For this purpose, we introduce the time dependence in (8) by taking its time derivative and then write the result as
\begin{equation}
\frac{\partial \rho_g}{\partial t}\,=\,\frac{\partial \rho_0}{\partial t}\,=\,- \frac{1}{4\pi G} \mathbf{\nabla}\cdot \frac{\partial \mathbf{E}_g}{\partial t}.
\end{equation}
Now using equation (17) in equation (14) we obtain 
\begin{equation}
\mathbf{\nabla}\cdot\left(\mathbf{j}_0\,-\,\frac{1}{4\pi G} \frac{\partial \mathbf{E}_g}{\partial t}\right)\,=\,0.
\end{equation}
The quantity inside the parenthesis of (18) is a vector whose divergence is zero. Since $\mathbf{\nabla}\cdot(\mathbf{\nabla}\times\mathbf{X})\,=\,0$\, for any vector $\mathbf{X}$, the vector inside the parenthesis of (18) can be expressed as the curl of some other vector, say $\mathbf{h}$. Mathematically speaking, the equation (18) admits of two independent solutions:
\begin{equation}
\mathbf{\nabla}\times\mathbf{h}\,=\,\pm \left(\mathbf{j}_0\,-\,\frac{1}{4\pi G}\frac{\partial \mathbf{E}_g}{\partial t}\right).
\end{equation} 
Except for the sign ambiguity in (19) (to be clear up soon), we now realize that we have just arrived at (in some form) the gravitational analogue of the Amp\`{e}re-Maxwell law in classical electrodynamics:
\begin{subequations}
\begin{align}
\mathbf{\nabla}\times\mathbf{H}\,=\,+\,\mathbf{j}_e\,+\,\epsilon_0 \frac{\partial \mathbf{E}}{\partial t}  \qquad \quad \mbox{or} \\
\mathbf{\nabla}\times\mathbf{B}\,=\,+\,\mu_0\mathbf{j}_e\,+\,\epsilon_0\mu_0 \frac{\partial \mathbf{E}}{\partial t}, 
\end{align}
\end{subequations}
\noindent
where $\mathbf{H}$ is called the magnetic field, $\mathbf{B} = \mu_0\mathbf{H}$ is called the magnetic induction field in vacuum, $\mathbf{E}$ represents the electric field, $\mathbf{j}_e\,=\,\rho_e \mathbf{v} =$ the electric charge current density ($\rho_e$ being electric charge density), $\epsilon_0$ and $\mu_0$ respectively represents the electric permitivity and the magnetic permeability of free space or vacuum and they are related to the speed of electromagnetic wave in vacuum, $c$ , by the relation
\begin{equation}
c\,=\,\frac{1}{\sqrt{\epsilon_0\mu_0}}
\end{equation}
and $c$ is a universal constant of nature. In fact, there are four field equations in electrodynamics which are collectively known as Maxwell's equations:
\begin{subequations}
\begin{align}
\mathbf{\nabla}\cdot \mathbf{E}\,=\,+\,\frac{\rho_e}{\epsilon_0}  \\
\mathbf{\nabla}\times\mathbf{B}\,=\,+\,\mu_0\mathbf{j}_e\,+\,\frac{1}{c^2} \frac{\partial \mathbf{E}}{\partial t} \\
\mathbf{\nabla}\cdot \mathbf{B}\,=\,0  \\
\mathbf{\nabla}\times\mathbf{E}\,=\, -\,\frac{\partial \mathbf{B}}{\partial t} 
\end{align}
\end{subequations}
\noindent
Maxwell's equations (22a-22d) form the basis of all classical electromagnetic phenomena including the production and transmission of electromagnetic waves carrying energy and momentum even in vacuum. When combined with the Lorentz force equation,
\begin{equation}
\mathbf{F}_L\,=\,q\mathbf{E}\,+\,q\mathbf{v}\times \mathbf{B}
\end{equation}
and Newton's 2nd law of motion (11),
these equations provide a complete description of classical dynamics of interacting charged particles and electromagnetic fields. Do analogous phenomena occur in gravitational physics? For instance,
Maxwell's equations predict the existence of electromagnetic waves that travel through vacuum (where $\rho_e = 0$ and $\mathbf{j}_e = \vec{0}$) at a universal speed $c = 3\times 10^8$ m/s and the wave equations for the fields $\mathbf{E}$ and $\mathbf{B}$ in vacuum are obtainable from (22a-22d) as
\begin{subequations}
\begin{align}
\mathbf{\nabla}^2\mathbf{E}\,-\,\frac{1}{c^2}\frac{\partial ^2 \mathbf{E}}{\partial t^2}\,=\,\mathbf{0},  \\
\mathbf{\nabla}^2\mathbf{B}\,-\,\frac{1}{c^2}\frac{\partial ^2 \mathbf{B}}{\partial t^2}\,=\,\mathbf{0}.
\end{align}
\end{subequations}
Before we explore analogous wave equations for the $\mathbf{E}_g$ and $\mathbf{h}$ fields, we note that one of the solutions in (19) would correspond to the reality and the other might be a mathematical possibility having no or new physical significance in which we are not presently interested in. We should note that Nature does not behave in two different ways at the same place and time - A physical quantity has to take a unique value at a particular place and time to be real. For this reason and for our present purpose, we have to choose one of the
two solutions that may correspond to the reality. Which one to choose? We can answer this question by following the rule of \emph{study by analogy}. A close look at Maxwell's in-homogeneous equations (22a, 22b) suggests us:\\
(a) that the source terms ($\rho_e $) and ($\vec{j}_e$) are of the same sign, \\
(b) that the analogue of $\epsilon_0$ in gravity is $\epsilon_{0g}$:
 \begin{equation}
\epsilon_{0g}\, =\, \frac{1}{4\pi G} \qquad \quad \left(\mbox{or} \quad \epsilon_{0g} = \,- \, \frac{1}{4\pi G}\right)    
\end{equation}
which we may call the gravito-electric permitivity of vacuum or whatever better name one may assign to it, and  \\
(c) to introduce a new constant $\mu_{0g}$ as 
\begin{equation}
\mu_{0g}\,=\,\frac{4\pi G}{c_{g} ^2} \qquad \quad \left(\mbox{or} \quad \mu_{0g}\,=\,-\,\frac{4\pi G}{c_{g} ^2}\right)
\end{equation}
which is the gravitational analogue of $\mu_0$ and $c_g$ is a new universal constant for vacuum having the dimension of velocity (which will turn out as the speed of gravitational waves in vacuum, if they exist) such that we get a relation an analogous to (21) as 
\begin{equation}
c_g\,=\,\frac{1}{\sqrt{\epsilon_{0g}\mu_{0g}}}.
\end{equation}
Accepting the suggestion (a), we are naturally led to choose the solution (19) having a negative sign before $\vec{j}_0$ as the real solution in view of equation (8), where the source term $\rho_0$ has a negative sign before it. Now, we multiply\footnote{We can multiply a positive scalar quantity with a vector equation without altering the physical content of that equation as we only re-scaling the those vectors in some new units.} $\mu_{0g}$ with our chosen equation (19) to get 
\begin{equation}
\mathbf{\nabla}\times\mathbf{B_{g}}\,=\,\mathbf{\nabla}\times(\mu_{0g}\mathbf{h})\,=\,-\,\mu_{0g}\mathbf{j}_0\,+\,\frac{1}{c_g^2}\frac{\partial \mathbf{E}_g}{\partial t},
\end{equation} 
\noindent
where we considered ($\mu_{0g}=4\pi G/c_g^2$) and defined $\mathbf{B}_{g} = \mu_{0g}\mathbf{h}$ as the gravitomagnetic induction field. Now taking the curl of (28) and using the vector identity $\mathbf{\nabla}\times (\mathbf{\nabla}\times \mathbf{B}) = \mathbf{\nabla}(\mathbf{\nabla}\cdot \mathbf{B}) - \mathbf{\nabla}^2\mathbf{B}$, we get
\begin{equation}
\mathbf{\nabla}(\mathbf{\nabla}\cdot \mathbf{B}_{g}) - \mathbf{\nabla}^2\mathbf{B}_{g} = -\mu_{0g}\mathbf{\nabla}\times \mathbf{j}_0 +\frac{1}{c_g^2}\frac{\partial}{\partial t}(\mathbf{\nabla}\times \mathbf{E}_g). 
\end{equation}
\noindent
In vacuum (where $\mathbf{j}_0 = \vec{0}$) equation (29) will reduce to a wave equation for the $\mathbf{B}_{g}$ field, viz.,
\begin{equation}
\mathbf{\nabla}^2\mathbf{B}_{g}\,-\,\frac{1}{c_g^2}\frac{\partial ^2 \mathbf{B}_{g}}{\partial t^2}\,=\,\vec{0}, 
\end{equation}
\noindent
provided the following two conditions:
\begin{equation}
\mathbf{\nabla}\cdot \mathbf{B}_{g}\,=\,0 \quad {\mbox{and}}
\end{equation}
\begin{equation}
\mathbf{\nabla}\times \mathbf{E}_g\,=\,-\,\frac{\partial \mathbf{B}_{g}}{\partial t}
\end{equation}
\noindent
are fulfilled. 
Now the taking curl of the equations (32) and using (28) we get the following equations for vacuum (where $\mathbf{j}_0 = 0$)
\begin{equation}
\mathbf{\nabla}(\mathbf{\nabla}\cdot \mathbf{E}_g) - \mathbf{\nabla}^2\mathbf{E}_g = -\,\frac{\partial}{\partial t}\mathbf{\nabla}\times \mathbf{B}_{g} = -\frac{1}{c_g^2}\frac{\partial ^2\mathbf{E}_g}{\partial t^2}.
\end{equation}
Since $\rho_0 = 0 =\mathbf{\nabla}\cdot \mathbf{E}_g$ in vacuum, this equation yields a wave equation for the $\mathbf{E}_g$ field:
 \begin{equation}
\mathbf{\nabla}^2\mathbf{E}_g = \,\frac{1}{c_g^2}\frac{\partial ^2\mathbf{E}_g}{\partial t^2}. \\
\end{equation}  
Thus, we notice that we have arrived at gravitational wave producing Gravito-Maxwell equations representing what call here as \\\\
{\bf Non-Relativistic Maxwellian Gravity (NRMG)}: 
\begin{subequations}
\begin{align}
\mathbf{\nabla}\cdot\mathbf{E}_g\,=\,-\rho_0/\epsilon_{0g} \\
\mathbf{\nabla}\times\mathbf{B}_g\,=\,-\,\mu_{0g}\mathbf{j}_0\,+\,\frac{1}{c_g^2}\frac{\partial \mathbf{E}_g}{\partial t}  \\
\mathbf{\nabla}\cdot\mathbf{B}_g\,=\,0 \\
\mathbf{\nabla}\times\mathbf{E}_g\,=\,-\,\frac{\partial \mathbf{B}_g}{\partial t}
\end{align}
\end{subequations}
\noindent
To complete the dynamical picture, we explore the gravitational the analogue of Lorentz force law (23) below adopting the logic and methods of Schwinger et al. \cite{2} in their derivation of the Lorentz force law. \\
Consider two inertial frames $S$ and $S^\prime$ having a relative velocity $\mathbf{v}$ between them. Let all the masses are in static arrangement in one of these frames, say $S^\prime$, which is moving with velocity $\mathbf{v}$. This way we introduce the time dependence of $\rho$ and $\mathbf{E}_g$ in the simplest way by assuming all masses are in uniform motion with a common velocity $\mathbf{v}$ with respect to $S$-frame. Here we use the Galileo-Newton principle of relativity (masses at rest and masses with a common velocity viewed by a co-moving observer are physically indistinguishable) and insist that physical laws are the same in the two inertial frames. Further, we assume that $|\mathbf{v}| << c_g = c$. To catch up with the moving masses one would have to move with velocity $\mathbf{v}$. Accordingly, the time derivative in the co-moving system, in which the masses are at rest, is the sum of explicit time dependent and co-ordinate dependent contributions,
 \begin{equation}
 \frac{d}{dt}\,=\,\frac{\partial}{\partial t}\,+\,\mathbf{v}\cdot \mathbf{\nabla}
\end{equation} 
so, in going from static system to uniformly moving system, we make the replacement 
\begin{equation}
\frac{\partial}{\partial t}\, \longrightarrow \,\frac{d}{dt}\,=\,\frac{\partial}{\partial t}\,+\,\mathbf{v}\cdot \mathbf{\nabla}.
\end{equation} 
The equation for constancy of gravitational field in equation (15c) becomes, in the moving system 
\begin{equation}
\mathbf{0} = \frac{\partial \mathbf{E}_g}{\partial t}\,\longrightarrow \,\mathbf{0}\,=\,\frac{d\mathbf{E}_g}{dt}\,=\,\frac{\partial \mathbf{E}_g}{\partial t}\,+\,(\mathbf{v}\cdot \mathbf{\nabla})\mathbf{E}_g.
\end{equation}
 The vector identity for constant $\mathbf{v}$ is 
\begin{equation}
\mathbf{\nabla}\times (\mathbf{v}\times \mathbf{E}_g)= \mathbf{v}(\mathbf{\nabla}\cdot\mathbf{E}_g) - (\mathbf{v}\cdot \mathbf{\nabla})\mathbf{E}_g.
\end{equation}
Now using Gauss's law of gravitostatics (35a) in the above identity we get 
\begin{equation}
\begin{split}
\mathbf{\nabla}\times (\mathbf{v}\times \mathbf{E}_g)&= -\frac{(\rho_0 \mathbf{v})}{\epsilon_{0g}}-(\mathbf{v}\cdot  \mathbf{\nabla})\mathbf{E}_g  \\
                                                     &= -\frac{\mathbf{j}_0}{\epsilon_{0g}}-(\mathbf{v}\cdot \mathbf{\nabla})\mathbf{E}_g.
\end{split}
\end{equation}
Again, from equations (38) and (40), we get 
\begin{equation}
\mathbf{\nabla}\times (\mathbf{v}\times \mathbf{E}_g)=\,-\,\frac{\mathbf{j}_0}{\epsilon_{0g}} + \frac{\partial \mathbf{E}_g}{\partial t}.
\end{equation}
Multiplication of equation (41) by $c_g^{-2}$ gives us 
\begin{equation}
\mathbf{\nabla}\times \left(\frac{\mathbf{v}\times \mathbf{E}_g}{c_g^2}\right)\,=\,-\,\mu_{0g}\mathbf{j}_0 + \frac{1}{c_g^2}\frac{\partial \mathbf{E}_g}{\partial t}.
\end{equation}
Equation (42) will agree with the equation (35b) of NRMG, only when 
\begin{equation}
\mathbf{B}_g\,=\,\frac{\mathbf{v}\times \mathbf{E}_g}{c_g^2}.
\end{equation}
Now consider the vector identity for constant $\mathbf{v}$ 
\begin{equation}
\mathbf{\nabla}\times (\mathbf{v}\times \mathbf{B}_g)= \mathbf{v}(\mathbf{\nabla}\cdot\mathbf{B}_g)\,-\,(\mathbf{v}\cdot \mathbf{\nabla})\mathbf{B}_g =\,-\,(\mathbf{v}\cdot \mathbf{\nabla})\mathbf{B}_g  
\end{equation}
where we have used $\mathbf{\nabla}\cdot\mathbf{B}_g\,=\,0$. Moreover, in the co-moving system where the masses are at rest - static - the $\mathbf{B}_g$ field should also not change with time:
\begin{equation}
\frac{d\mathbf{B}_g}{dt}\,=\,\frac{\partial \mathbf{B}_g}{\partial t}\,+\,(\mathbf{v}\cdot \mathbf{\nabla})\mathbf{B}_g\,=\,\mathbf{0}.
\end{equation}
From (44)and (45) we get 
\begin{equation}
\frac{\partial \mathbf{B}_g}{\partial t}\,=\,\mathbf{\nabla}\times (\mathbf{v}\times \mathbf{B}_g).
\end{equation} 
Again, the vector identity
\begin{equation}
\mathbf{\nabla}^2 \mathbf{E}_g\,= \,\mathbf{E}_g(\mathbf{\nabla}\cdot \mathbf{E}_g)\,-\,\mathbf{\nabla}\times (\mathbf{\nabla}\times \mathbf{E}_g)
\end{equation}
 gives us 
the left hand side of the wave equation (34) for $\mathbf{E}_g$ in vacuum (i.e. outside the mass distribution where $\mathbf{\nabla}\cdot \mathbf{E}_g\,=\,0$) as
\begin{equation}
\mathbf{\nabla}^2 \mathbf{E}_g = - \mathbf{\nabla}\times(\mathbf{\nabla}\times 
\mathbf{E}_g).
\end{equation}
By means of the equation (35b) of NRMG and equation (46), the right side of the wave equation (34) becomes ($\mathbf{j}_0 = \mathbf{0}$ outside the charge distribution)
\begin{equation}
\begin{split}
\frac{1}{c_g^2}\frac{\partial ^2 \mathbf{E}_g}{\partial t^2}&= + \frac{\partial }{\partial t}(\mathbf{\nabla}\times \mathbf{B}_g)  \\
                                                     &= + \mathbf{\nabla}\times \left[\mathbf{\nabla}\times (\mathbf{v}\times \mathbf{B}_g)\right].
\end{split}
\end{equation}
Equations (48-49) show that the wave equation for $\mathbf{E}_g$ (34) will hold if 
\begin{equation}
\mathbf{E}_g\,=\,-\,\mathbf{v}\times \mathbf{B}_g.
\end{equation}
\noindent
But this cannot be completely correct since as 
$\mathbf{v}\rightarrow \mathbf{0} \Rightarrow \mathbf{E}_g \rightarrow \mathbf{0}$. No gravitostatics ! However, all that is necessary is that the curl of this $\mathbf{E}_g$ in (50)  should be valid:
\begin{equation}
\mathbf{\nabla}\times \mathbf{E}_g\,=\,-\,\mathbf{\nabla}\times (\mathbf{v}\times \mathbf{B}_g)
\end{equation}
\noindent
or, if we use equation (46),
\begin{equation}
\mathbf{\nabla}\times \mathbf{E}_g\,=\,-\,\frac{\partial \mathbf{B}_g}{\partial t}.
\end{equation}
This is consistent with gravitostatics since it generalizes $\mathbf{\nabla}\times \mathbf{E}_g\,=\,\mathbf{0}$ to time-dependent situation. \\
\noindent
Now we are ready to address the question: What replaces the equation 
$\mathbf{F}\,=\,m_0\mathbf{E}_g$ to describe the force on a point particle of mass $m_0$, when that particle moves with some non-relativistic velocity $\mathbf{v}$ in given $\mathbf{E}_g$ and $\mathbf{B}_g$ field? For this purpose, consider two coordinate systems, one in which the particle is at rest (co-moving coordinate system) and one in which it moves at velocity $\mathbf{v}$. Suppose, in the later coordinate system, the gravitational (or gravito-electric) and gravitomagnetic fields are given by $\mathbf{E}_g$ and $\mathbf{B}_g$, respectively. 
In the co-moving frame, the force on the particle is  
\begin{equation}
\mathbf{F}_g\,=\,m_0\mathbf{E}_g^{\mbox{eff}},
\end{equation}
\noindent 
where $\mathbf{E}_g^{\mbox{eff}}$ is the gravitational (or gravito-electric) field in this frame. In transforming to the co-moving frame, all the other masses - those responsible for $\mathbf{E}_g$ and $\mathbf{B}_g$ - have been given an additional counter velocity $-\mathbf{v}$. From equation (50), we then infer that ($+\mathbf{v}\times\mathbf{B}_g$) has the character of an additional gravito-electric field in the co-moving frame. Hence, the suggested $\mathbf{E}_g^{\mbox{eff}}$ is 
\begin{equation}
\mathbf{E}_g^{\mbox{eff}}\,=\,\mathbf{E}_g\,+\,\mathbf{v}\times\mathbf{B}_g, 
\end{equation}
\noindent
leading to the gravitational analogue of the Lorentz force law that we call Gravito-Lorentz force law: 
\begin{equation}
\mathbf{F}_{\mbox{gL}}\,=\, m_0\left(\mathbf{E}_g\,+\,\mathbf{v}\times\mathbf{B}_g\right). 
\end{equation}
This gravito-Lorentz force law when used in Newton's 2nd law of motion, 
\begin{equation}
m_0\frac{d\mathbf{v}}{dt} = \mathbf{F}_{\mbox{gL}},
\end{equation}
we get the following equation of motion of NRMG
\begin{equation}
\frac{d\mathbf{v}}{dt}\,=\, \mathbf{E}_g\,+\,\mathbf{v}\times\mathbf{B}_g.
\end{equation}
\noindent
Since $\mathbf{\nabla}\cdot \mathbf{B}_g\,=\,0$, $\mathbf{B}_g$ can be defined as the curl of some vector function, say $\mathbf{A}_g$:
\begin{equation}
\mathbf{B}_g\,=\,\mathbf{\nabla}\times \mathbf{A}_g,
\end{equation}
where $\mathbf{A}_g$ is the vector potential for NRMG. Using equation (58) in the Gravito-Faraday law of NRMG (35d), we get the following expression for the $\mathbf{E}_g$ in terms of $\phi_g$ and $\mathbf{A}_g$ as 
\begin{equation}
\mathbf{E}_g\,=\,-\,\mathbf{\nabla}\phi_g\,-\,\frac{\partial \mathbf{A}_g}{\partial t}.
\end{equation}
We now recognize that the equations of NRMG correspond to the equations (1-7) of GRMG-UG provided we set $c_g = c$. 
In this derivation, $c_g$ is an undetermined parameter of the theory, whose value may be obtained from some more advanced theory or from experiments. \\
Substituting the expression for $\mathbf{E}_g$ given by equation (59) and the expression for $\mathbf{B}_g$ defined by (58) in the in-homogeneous field equations (35a, 35b) of NRMG, we get the following expressions for the in-homogeneous equations (35a, 35b) in terms of potentials $(\phi_g, \mathbf{A}_g)$ as
\begin{equation}
\mathbf{\nabla}^2\phi_g - \frac{1}{c_g^2} \frac{\partial ^2\phi_g}{\partial t^2}=
\frac{\rho_0}{\epsilon_{0g}}, 
\end{equation}
\begin{equation}
\mathbf{\nabla}^2\mathbf{A}_g - \frac{1}{c_g^2} \frac{\partial ^2\mathbf{A}_g}{\partial t^2}
=\mu_{0g}\mathbf{j}_0, 
\end{equation}
\noindent
if the following gravitational Lorenz gauge condition, 
\begin{equation}
\mathbf{\nabla}\cdot\mathbf{A}_g\,+\,\frac{1}{c_g^2}\frac{\partial \phi_g}{\partial t}\,=\,0,
\end{equation}
is imposed. These will determine the generation of gravitational waves by prescribed gravitational mass and mass current distributions. Particular solutions of (60) and (61) in vacuum are
\begin{equation}
\phi_g(\mathbf{r}, t) = -\frac{1}{4\pi \epsilon_{0g}}\int \frac{\rho_0(\mathbf{r}^\prime , t^\prime)}{|\mathbf{r} -\mathbf{r}^\prime|}dv^\prime \quad {\mbox{and}}
\end{equation}
\begin{equation}
\mathbf{A}_g(\mathbf{r}, t) = -\frac{\mu_{0g}}{4\pi }\int \frac{\mathbf{j}_0(\mathbf{r}^\prime , t^\prime)}{|\mathbf{r} -\mathbf{r}^\prime|}dv^\prime,
\end{equation}
\noindent
where $t^\prime = t - |\mathbf{r} - \mathbf{r}^\prime|/c_g$ is the retarded time and $dv^\prime$ is an elementary volume element at $\mathbf{r}^\prime$. Thus, we saw that retardation in gravity is possible in Newtonian space and time in the same procedure as we adopt in  electrodynamics. Hence, we have reasons to strongly disagree with Rohrlich's conclusion \cite{19}: ``{\it Because the Newtonian theory is entirely static, retardation is not possible until the correction due to deviations from Minkowski space is considered"}.
 According to the present field theoretical view, gravitation, like electromagnetism and all other fundamental interactions, acts locally through fields: A mass at one point produces a field, and this field acts on whatever masses with which it comes into contact \cite{20}. Because of the finite propagation speed of the fields, gravitational effects/information propagate at finite speed, that is the cause behind retarded interaction. However, in case of static or quasi-static mass distributions, retardation effects are negligible and hence no distinction can be made between local interaction and action-at-a-distance. By introducing the concept of physical fields carring energy and momentum\footnote{Field is the material Newton was serching for: the field is material because it possesses an energy density \cite{20}.}, one can address Newton's ``action-at-a-distance'' problem within Newton's world of physics by extending his field equations to time-dependent fields, sources and searching for the conditions for the existence of gravitational waves in free space traveling at a finite speed. This is what Heaviside had done in 1893 \cite{4,5}. May be Einstein had not seen Heaviside's field equations when he was working on his relativistic theory of gravity. Had Einstein seen Heaviside's field equations, his remark on Newton's theory of gravity would have been different than what he made before the 1913 congress of natural scientists in Vienna \cite{21},viz.,  
\begin{quotation}
After the un-tenability of the theory of action at distance had thus been proved
in the domain of electrodynamics, confidence in the correctness of Newton's
action-at-a-distance theory of gravitation was shaken. One had to believe
that Newton's law of gravity could not embrace the phenomena of gravity in
their entirety, any more than Coulomb's law of electrostatics embraced the
theory of electromagnetic processes.
\end{quotation}
\section{Special Relativistic Maxwellian Gravity (SRMG)}
\label{sec:3}
With the establishment of special relativity (SR) theory and the equivalence of mass and
energy, the meaning of the inertial mass and gravitational mass
became ambiguous, because SR suggests two inertial mass-energy
concepts: (1) the Lorentz invariant rest-mass $m_0\,=\,E_0/c^2$ ($E_0$ = rest-energy, which is the sum
total of all forms of energy in the rest frame of a body or
particle) and (2) the mass
attributed to the relativistic energy $m\,= \,E/c^2$ ($E\,=$
sum of all forms of energy at rest and motion) which is not Lorentz-invariant. The qualitative
distinction that existed between two inertial mass concepts in
Newtonian mechanics became quantitatively distinct and clear in
SR. Now, one fundamental question arises, ``What form of mass
(or energy) should be the source of gravity ($m_g$) in a
relativistic version of Newtonian gravity?" In any construction of
a field theory of gravity compatible with SR and the
correspondence principle by which a relativistic theory gravity is
reducible to Newtonian gravity, a decision on which form of
``mass" (or energy) is the source of gravity has to be taken. Such
a decision, as Price \cite{22} has rightly pointed out, will be
crucial not only to the resolution of the ambiguity mentioned
above but also to the issue of the nonlinear nature of
gravity. One of the Eddington's \cite{23} four reasons to feel dissatisfied with Newton's Law of 
gravitation is appropriate here to quote:
\begin{quote}
The most serious objection against the Newtonian Law as an exact law was that
it had become ambiguous. The law refers to the product of the masses of the two bodies; but the mass depends on the velocity- a fact unknown in Newton's days. Are we to take the variable mass, or the mass reduced to rest? Perhaps a learned judge, interpreting Newton's statement like a last will and testament, could give a decision; but that is scarcely the way to settle an important point in scientific theory.
\end{quote} 
In his construction of a relativistic theory of gravity popularly known as General Relativity (GR), Einstein has taken a decision in favor of the equality of $m$ with $m_g$. For a theoretical justification of this decision,
Einstein by writing Newton's equation of motion in a gravitational field (in our present mathematical notation) as
\begin{equation}
m\frac{d\mathbf{v}}{dt}\,=\,m_g \mathbf{E}_g
\end{equation}
(wrongly!) inferred from it \cite{24}:
\begin{quote}
It is only when there is numerical equality between the inertial and gravitational mass that the acceleration is independent of the nature of the body.
\end{quote}
This inference is often expressed in one of the two ways:\\
(S1) that the particle's motion is mass independent, or \\
(S2) that the particle's inertial mass $m$ = its
gravitational mass $m_g$.\\
The two statements (S1) and (S2) are sometimes used
interchangeably as the {\it weak equivalence principle} (WEP) in
the literature \cite {25,26,27}. This use of terminology is rather
confusing, as the two statements are logically independent
\cite{28}. They happen to coincide in the context of
Galileo-Newtonian physics where $m_0 = m = m_g$ but may diverge in the context of
special relativity where $m \neq m_0$ and Einstein's wrong inference of $m_0\neq m = m_g$ from a non-relativistic equation (65), which is 
exactly the equation (13) of NRMG, where $m = m_0$ and $m_g = m_0$ is a condition for Galileo's law of 
Universality of Free Fall to be true. 
To explore this possibility, to get new insights for
making Newtonian gravity compatible with the SR, to regard old
problems from a new angle, we re-examined \cite{3} an often cited
\cite{29,30} Salisbury-Menzel's \cite{31} thought experiment
(SMTE) from a new perspective as discussed in the following subsection. 
Before that the author would like to remark that perhaps Einstein, himself, was not satisfied with his above inference of $m_g = m$, as we can sense from  his another statement on the equality of $m_g$ with $m$ \cite{32,33}:
\begin{quote}
The proportionality between the inertial and gravitational masses holds for all bodies without 
exception, with the (experimental) accuracy achieved thus far, so that we may assume its general validity until proved otherwise.
\end{quote}  
The last three words of Einstein's above statement,`{\it until proved otherwise}', show that he was very cautious and not very confident of what he was stating. Based on the experimental results available up to 1993, Mashhoon \cite{30} noted that the observational evidence for the principle of equivalence of gravitational and inertial masses is not yet precise enough to reflect the wave nature of matter and radiation in their interactions with gravity (see other references on equivalence principle in \cite{3,30}).   
\subsection{Re-Examination of SMTE to Show $m_0 = m_g$}
Consider a system of two non-spinning point-like charged particles
with charges $q_1$ and $q_2$ with respective rest masses $m_{01}\, (= E_{01}/c^2)$
and $m_{02}\, (= E_{02}/c^2)$ such that they are at rest in an inertial frame
$S^\prime$ under equilibrium condition due to a mutual balance of
the force of Coulombic repulsion ($\mathbf{F}_C^\prime$) and the Newtonian 
gravitostatic attraction ($\mathbf{F}_N^\prime$) between them. Our aim is to 
investigate the condition of equilibrium of this two-particle system (realizable 
in a Laboratory by taking two perfectly identical spherical metallic spheres
having requisite masses and charges so that they are in
equilibrium) and in different inertial frames in relative motion.
For our re-examination purpose, suppose that the particles are positively charged and
they are in empty space. Let the particle No.2 be positioned at
the origin of $S^\prime$-frame and $\mathbf{r}^\prime$ be the position
vector of the particle No.1 with respect to the particle No.2. In
this $S^\prime$-frame the condition of equilibrium is fulfilled by
\begin{equation}
\mathbf{F}_C^\prime\,+\mathbf{F}_N^\prime\,=\,\frac{q_1q_2 \vec
r^\prime}{4\pi \epsilon_0 {r^\prime}^3}\,-\,\frac{Gm_{01}m_{02}\vec
r^\prime}{{r^\prime}^3} = {\bf 0}
\end{equation}
where $r^\prime\,=\,|\vec r^\prime|$ and other symbols have their usual meanings. From (66) we get
\begin{equation}
\frac{q_1q_2}{4\pi \epsilon_0}=G
m_{01}m_{02}=\frac{m_{01}m_{02}}{4\pi \epsilon
_{0g}}\qquad(\epsilon_{0g}=1/4\pi G)
\end{equation}
Equation (67) represents the condition of equilibrium, in terms of
the charges and rest masses (or rest energies) of the particles, under which an
equilibrium can be ensured in the $S^\prime$-frame. For example,
if each metallic sphere is given a charge of $1\times 10^{-6}$
Coulomb, then the rest mass of each sphere should be $1.162\times
10^{4}$ kg, to fulfill the equilibrium condition (67) in a laboratory experiment.  \\ 
Now, let us investigate the problem of equilibrium of the said
particle system from the point of view of an observer in another
inertial frame $S$, in uniform relative motion with respect to the
$S^\prime$-frame. To simplify the investigation, let the relative
velocity $\mathbf{v}$ of $S$ and $S^\prime$-frame be along a common
$X/X^\prime$-axis with corresponding planes parallel as usual.
Since the particles are at rest in $S^\prime$-frame, both of them
have the same uniform velocity $\mathbf{v}$ relative to the $S$-frame.
Let the position vector of the particle No.1 with respect to the
particle No.2 as observed in the $S$-frame be $\mathbf{r}$ and the angle
between $\mathbf{v}$ and $\mathbf{r}$ be $\theta$.\\
For an observer in the $S$-frame, the force of electric origin on
either particle (say on particle No.1 due to particle No.2) is no
more simply a Coulomb force, but a Lorentz force, viz.,
\begin{equation}
\mathbf{F}_L\,=\,q_1\mathbf{E}_2\,+\,q_1\mathbf{v}\times \mathbf{B}_2
\end{equation}
where
\begin{equation}
\mathbf{E}_2\,=\,\frac{q_2(1\,-\,\beta^2)\mathbf{r}}{4\pi\epsilon_0
r^3\left(1\,-\,\beta ^2
\sin^2{\theta}\right)^{3/2}},\qquad(\mathbf{\beta}  = \mathbf{v}/c)
\end{equation}
\begin{align}
\mathbf{B}_2\,&= \frac{\mathbf{v}\times \mathbf{E}_2}{c^2} = \frac{(q_2\mathbf{v})\times \mathbf{r}\,(1\,-\,\beta^2)}{4\pi \epsilon_0 c^2\,r^3\left(1\,-\,\beta ^2 \sin^2{\theta}\right)^{3/2}} \nonumber \\
         &= \frac{\mu_0}{4\pi}\,\frac{(q_2\mathbf{v})\times \mathbf{r}\,(1\,-\,\beta^2)}{r^3\left(1\,-\,\beta ^2
\sin^2{\theta}\right)^{3/2}}
\end{align}
\begin{equation}
\mathbf{r}\,=\,\frac{\mathbf{r}^\prime \left(1\,-\,\beta
^2\sin^2{\theta}\right)^{1/2}}{\left(1\,-\,\beta ^2\right)^{1/2}}.
\end{equation}
What about the force of gravitational interaction as observed in
the $S$-frame? It can not simply be a Newtonian force but
something else, otherwise the particle system will not remain in
equilibrium in the $S$-frame. Such a situation will amount to a
violation of the principle of relativity in special relativity. A
null force should remain null in all inertial frames. Therefore, a
new force law of gravity has to be invoked so that the equilibrium
is maintained in accordance with the principle of relativity (Lorentz invariance of physical laws). 
Let this new unknown force be represented by $\mathbf{F}_{gL}$ such that
the equilibrium condition in $S$-frame is satisfied as:
\begin{equation}
\mathbf{F}_{gL}\,+\,\mathbf{F}_L\,=\,\mathbf{0}\qquad\implies \, \mathbf{F}_{gL}\,= -\,\mathbf{F}_L.
\end{equation}
Taking into account the equations (68)-(71), $\mathbf{F}_{gL}$ in
equation (72) can be expressed as:
\begin{align}
\vec F_{gL}\,&=\,-\,\frac{q_1q_2\left(1\,-\,\beta^2\right)\vec
r}{4\pi \epsilon _0 r^3\left(1\,-\,\beta ^2
\sin^2{\theta}\right)^{3/2}} \nonumber \\
         &\quad {} -\,\frac{\mu_0}{4\pi}\,\frac{q_1q_2\vec
v \times (\vec v\times \vec r)\left(1\,-\,\beta^2\right)}{
r^3\left(1\,-\,\beta ^2 \sin^2{\theta}\right)^{3/2}}.
\end{align}
Now, using equation (67), we can eliminate $q_1q_2$ from equation
(73) to get the expression for $\vec F_{gL}$ in terms of $m_{01},
m_{02}$ and $G$ as:
\begin{align}
\vec F_{gL}\,&=\,-\,\frac{Gm_{01}m_{02}\left(1\,-\,\beta^2\right)\vec
r}{r^3\left(1\,-\,\beta ^2
\sin^2{\theta}\right)^{3/2}} \nonumber \\
         &\quad {} -\,\frac{G}{c^2}\,\frac{m_{01}m_{02}\vec
v \times (\vec v\times \vec r)\left(1\,-\,\beta^2\right)}{
r^3\left(1\,-\,\beta ^2 \sin^2{\theta}\right)^{3/2}} \nonumber \\ 
           &= \,-\,\frac{1}{4\pi \epsilon_{0g}}\frac{m_{01}m_{02}\left(1\,-\,\beta^2\right)\vec r}{r^3\left(1\,-\,\beta ^2
\sin^2{\theta}\right)^{3/2}} \nonumber \\
         &\quad {} -\,\frac{\mu_{0g}}{4\pi}\,\frac{m_{01}m_{02}\vec
v \times (\vec v\times \vec r)\left(1\,-\,\beta^2\right)}{
r^3\left(1\,-\,\beta ^2 \sin^2{\theta}\right)^{3/2}}, 
\end{align}
\noindent
where
\begin{equation}
\epsilon_{0g}\,=\,\frac{1}{4\pi G},\,\,
\mu_{0g}\,=\,\frac{4\pi G}{c^2} \implies c = \frac{1}{\sqrt{\epsilon_{0g}\mu_{0g}}}. 
\end{equation}
By comparing the quantities in equation (75) with those in equations (25-27), we immediately find that 
\begin{equation}
c_g\,=\,\frac{1}{\sqrt{\epsilon_{0g}\mu_{0g}}}\,=\,c.
\end{equation}
Now, (74) may be rearranged to the following form to represent the Gravito-Lorentz force law of SRMG:
\begin{equation}
\vec F_{gL}= m_{01}\vec E_{g2}+m_{01}\vec v \times \vec
B_{g2} 
\end{equation}
where 
\begin{align}
\vec E_{g2}\,&= \,-\,\frac{1}{4\pi \epsilon_{0g}}\,\frac{m_{02}(1\,-\,\beta^2)\vec r}{r^3\left(1\,-\,\beta^2\sin^2\theta\right)^{3/2}} \nonumber \\
          &\simeq  -\,\frac{1}{4\pi \epsilon_{0g}}\,\frac{m_{02}\mathbf{r}}{r^3} 
          \qquad (\mbox{when}\, \beta << 1),
\end{align}
\begin{align}
\vec B_{g2}\,&= \frac{\vec v\times \vec
E_{g2}}{c^2}\,=\,-\,\frac{\mu_{0g}}{4\pi}\,\frac{(m_{02}\vec v)\times \vec
r\,(1\,-\,\beta^2)}{r^3\left(1\,-\,\beta^2\sin^2\theta\right)^{3/2}} \nonumber \\
          &\simeq  -\,\frac{\mu_{0g}}{4\pi}\,\frac{(m_{02}\vec v)\times \vec
r}{r^3} \qquad (\mbox{when}\, \beta << 1).
\end{align}
Equations (77-79) are in complete formal analogy with the equations (68-70) of classical electromagnetism in its relativistic version. Thus, from the requirement of the frame-independence of the equilibrium condition, we not only obtained a gravitational analogue of the Lorentz-force law expressed by equation (77) but also unexpectedly found the Lorentz-invariant rest mass as the gravitational analogue of the electric charge by electromagnetic analogy. From this analysis, the gravitational charge (or rest mass) invariance may be interpreted as a consequence of the Lorentz-invariance of the physical laws. These findings are in conformity with Poinca\`{r}e's \cite{34} remark that {\it if equilibrium is to be a frame-independent condition, it is necessary for all forces of non-electromagnetic origin to have precisely the same transformation law as that of the Lorentz-force}. 
\noindent
Now, following Rosser's \cite{35} approach to classical electromagnetism via relativity one can obtain the field equations of SRMG as represented in equations (35a-35d) with $c_g = c$, from the equations (77-79). Alternatively, after recognizing our new findings from the above thought experiment, especially $m_g = m_0$ and $c_g = c$, one may follow the procedure we followed in NRMG to arrive at desired gravitational wave producing field equations. It is to be noted that the Lorrain's (see Lorrain in \cite{31}) exact special relativistic derivation of gravitational analogue of the magnetic force from SMTE matches with our SRMG results.  
\subsection{Lorentz co-variant formulation of SRMG} 
In Lorentz co-variant formulation, by introducing the space-time 4-vector $x^\alpha = (ct, \vec{x})$, proper (or rest) mass current density 4-vector $j^\alpha = (\rho_o c,\, \vec{j}_0)$ and second-rank anti-symmetric gravitational field strength tensor $f_{\alpha \beta}$:
\begin{equation}
f_{\alpha \beta}\,= \partial_\alpha A_\beta - \partial_\beta A_\alpha =
\begin{pmatrix}
  0         &    E_{gx}/c    &    E_{gy}/c   &   E_{gz}/c  \\
-E_{gx}/c   &    0       &   -B_{gz}         &   B_{gy}  \\ 
-E_{gy}/c   &    B_{gz}  &    0              &  -B_{gx}  \\ 
-E_{gz}/c   &   -B_{gy}  &    B_{gx}         &   0       
\end{pmatrix},
\end{equation}
one can rewrite the field equations of SRMG as:
\begin{equation}
\partial^\beta f_{\alpha \beta}\,= \partial^\beta\left(\partial_\alpha A_\beta - \partial_\beta A_\alpha\right)=\,\frac{4\pi G}{c^2}j_\alpha = \,\mu_{0g}j_\alpha,
\end{equation}
\begin{equation}
\partial_\alpha f_{\beta \gamma} +\partial_\beta f_{\gamma \delta} + \partial_\gamma f_{\alpha \beta} = 0,
\end{equation}
where $\alpha,\, \beta,\,\gamma$ are any three of the integers 0, 1, 2, 3; $j_\alpha = (\rho_oc,\,- \mathbf{j}_0)$ and $j^\alpha = (\rho_oc,\,\mathbf{j}_0)$; Gravito-Lorenz condition: $\partial^\alpha A_\alpha = 0$;\, $A_\alpha = (\phi_g,\,- \mathbf{A}_g)$ and $A^\alpha = (\phi_g,\,\mathbf{A}_g)$ with $\phi_g$ = scalar potential and $\mathbf{A}_g$ = vector potential; $\partial_\alpha \equiv \left(\partial/{c\partial t},\,\mathbf{\nabla}\right) \& \, \partial^\alpha \equiv \left(\partial/{c\partial t},\,-\mathbf{\nabla}\right)$. In this convention, the relativistic gravito-Lorentz force law takes the following form 
\begin{equation}
\frac{d^2x^\alpha}{d\tau^2} = f^{\alpha \beta}\frac{dx_\beta}{d\tau}
\end{equation}
where $\tau$ is the proper time along the particle's world-line and $f^{\alpha \beta}$ is given by 
\begin{equation}
f^{\alpha \beta} =\eta^{\alpha \gamma}f_{\gamma \delta}\eta^{\delta \beta} = 
\begin{pmatrix}
  0        &   -E_{gx}/c   &   -E_{gy}/c   &  -E_{gz}/c \\
E_{gx}/c   &    0          &   -B_{gz}     &   B_{gy}  \\ 
E_{gy}/c   &    B_{gz}     &    0          &  -B_{gx}  \\ 
E_{gz}/c   &   -B_{gy}     &    B_{gx}     &   0
\end{pmatrix}
\end{equation}
where the flat space-time metric tensor $\eta_{\alpha \beta} = \eta^{\alpha \beta}$ is represented by symmetric diagonal matrix with  
\begin{equation}
\eta_{00} = 1,\quad \eta_{11} = \eta_{22} = \eta_{33} = -1. 
\end{equation}
The relativistic equation of motion (83) is independent of the mass of the particle moving in an external gravito-electromagnetic (GEM) field $f^{\alpha \beta}$. Thus we saw that the motion of a particle in an external GEM field can be independent of its mass without any postulation on the equality of gravitational mass with frame-dependent inertial mass. Equation (83) is the relativistic generalization of Galileo's law of Universality of Free Fall (UFF) expressed through the non-relativistic equations of motion (13) and (57) and known to be true both theoretically and experimentally since Galileo's time.  \\
Now, if we introduce the energy momentum four vector:
\begin{equation}
p^\alpha = (p_0,\,\mathbf{p}) = m (U_0,\,\mathbf{U})
\end{equation}
where $p_0 = E/c$ and $U^\alpha$ is the 4-velocity, then with this $p^\alpha$ we can re-write equation (83) as 
\begin{equation}
\frac{dp^\alpha}{d\tau} = f^{\alpha \beta}p_\beta
\end{equation}
Thus, in SRMG the fields $f_{\alpha \beta}$ couple to the energy-momentum 4-vector of all particles of whatever rest masses they have, provided $m_g = m_0$ holds exactly. It is to be noted that the equation of motion (83 or 87) holds only in an inertial frame. Appropriate modifications are necessary for its application in non-inertial frames, as is done in non-relativistic physics by introducing pseudo-forces. 
\subsection{Original analysis of SMTE with assumption of $m_g = m = m_0/\sqrt{1 - v^2/c^2}$}
In the original analysis of SMTE \cite{31} Salisbury and Menzel (SM) axiomatically used flat space-time and assumed $m_g = m = m_0/\sqrt{1 - v^2/c^2}$ for their thought experimental demonstration of gravito-magnetic field (they called it {\it Gyron field}) and the gravitational analogue of Lorentz force law. From the analysis of their results, one can find that in the slow motion approximation, if the gravito-Lorentz force law is written in the following form
\begin{equation}
\mathbf{F}_{gL}^{SM} = m_0\frac{d\mathbf{v}}{dt} = m_0 \mathbf{E}_g + m_0\mathbf{v}\times \mathbf{B}_g, \quad \text{then} 
\end{equation} 
\begin{equation}
\mu_{0g}^{SM} = \frac{8\pi G}{c^2} \quad \text{while} \quad \epsilon_{0g}^{SM} = \frac{1}{4\pi G},
\end{equation}
which yields 
\begin{equation}
c_g^{SM} = \left(\mu_{0g}^{SM}\mu_{0g}^{SM}\right)^{-1/2} = c/\sqrt{2}.
\end{equation} 
On the other hand if one considers $c_g^{SM} = c$, then equation (88) has to be written in the following form: 
\begin{equation}
\mathbf{F}_{gL}^{SM} = m_0\frac{d\mathbf{v}}{dt} = m_0 \mathbf{E}_g + 2m_0\mathbf{v}\times \mathbf{B}_g. 
\end{equation}
We designate this type of gravity as linearized version of non-linear SRMG (SRGM-N) in flat space-time. The origins of the non-linearity of (SRGM-N), the appearance of the spurious value of $c_g = c/\sqrt{2}$ or a factor of ``2" in the gravitomagnetic force term (due to a supposed value of $c_g = c$) are all now traced to the adoption of Einstein's doubtful postulate on the equality of gravitational mass with the velocity dependent inertial mass. As such, SRMG-N does not correspond either to the NRMG when $c_g = c$ or to the non-relativistic limit of SRMG. 
\section{Discussions} 
\label{sec:4}
Several authors have suggested or obtained different Maxwell-type equations for gravity following 
different approaches. Some without using the formalism of GR \cite{36,37,38,39,40,41,42,43,44,45,46,47,48,49} and some using the formalism of GR in 
the weak field and linearized approximations. These are discussed in two separate sections below. 
The two subsequent sections concern discussions on Spin-1 Vector Gravity vs Spin-2 Tensor Gravity and  remarks on little known Heaviside's work on gravity.
\subsection{Maxwellian Gravity of Others without GR}
 Sciama\cite{36}, in 1953, hypothetically adopted SRMG (assuming $m_g = m_0$) to explain 
 the origin of inertia, calling it a toy model theory of gravity which differs from GR principally in three respects: 
 (a) It enables the amount of matter in the universe to be estimated from a knowledge of the gravitational constant, 
 (b) The principle of equivalence is a consequence of the theory, not an initial axiom and 
 (c) It implies that gravitation must be attractive. However, he concluded his paper mentioning 
 three limitations of such a theory: (i) It is incomplete because the relativistic form of Newton's law must 
 be derived from a tensor potential\footnote{This thought comes to anyone who believes in 
 $m_g = E/c^2$, where $E$ is the relativistic energy, which may not be true as per our findings discussed here.}, 
 not from a vector potential, (ii) It is difficult to give a consistent relativistic discussion of 
 the structure of the universe as a whole\footnote{An interesting topic of research not yet fully explored. As matter of scientific curiosity, one may explore the Universe from the new perspective of a vector gravitational theory.} and (iii) It is also difficult to
  describe the motion of light in a gravitational field\footnote{It is a deep question involving the 
  interaction of two fundamental fields which is beyond the scope of this paper but needs further investigation.}. 
  Carstoiu \cite{37,38}, in 1969, rediscovered Heaviside's gravitational equations (35a-35d) 
  (in our present notation as per the report of Brilloiun \cite{38}) assuming the existence of a 
  second gravitational field called {\it gravitational vortex} (here called gravito-magnetic field) 
  and assumed $c_g = c$ by electromagnetic analogy \cite{38}. In 1980, Cattani \cite{39} considered 
  linear equations for the gravitational field by introducing a new field by calling it the 
  {\it Heavisidian field} which depends on the velocities of gravitational charges in the same way 
  as a magnetic field depends on the velocities of electric charges and shown that a gravitational 
  field may be written with linear co-variant equations in the same way as for the electromagnetic field. 
  Cattani's equations differ from some important formulae of general relativity such as the gravitational
   radiation, Coriolis force by a factor of 4. In 1982, Singh \cite{40} considered a vector gravitational 
   theory having formal symmetry with the electromagnetic theory and explained the (a) precession of the
    perihelion of a planet (b) bending of light in the gravitational field and (c) gravitational red-shift 
    by postulating the self-interaction between a particle velocity and its vector potential. In 2004, 
    Flanders and Japaridze \cite{41} axiomatically used the field equations of SRMG (albeit without reference to \cite{3})
     and special relativity to explain the photon deflection and perihelion advance of Mercury in the 
     gravitational field of the Sun. Borodikhin \cite{42} explained the perihelion advance of Mercury, 
     gravitational deflection of light as well as Shapiro time delay by postulating a vector theory of 
     gravity in flat space-time that is nothing but SRMG. Borodikhin also showed that in a vector 
     theory of gravity, there exists a model for an expanding Universe. Jefimenko \cite{43,44} also
      deduced the equations of NRMG by extending Newton's gravitational theory to time-dependent sources 
      and fields and using the causality principle. Jefimenko assumed $c_g = c$ and  postulated a 
      gravito-Lorentz force. Recently, Heras \cite{45}, by recognizing the general validity of the 
      axiomatic approach to Maxwell's equations of electromagnetic theory, used those axioms to
       derive only the field equations (leaving out gravito-Lorentz force law) of SRMG, where the 
       in-variance of gravitational charge is considered. Other recent derivations of SRMG equations 
       from different approaches include the works of Nyambuya \cite{46}, Sattinger \cite{47}, 
       Vieira and Brentan \cite{48}. The historical objections of several researchers, starting from J. C. Maxwell \cite{3} upto Misner, Thorne and Wheeler (MTW, Sec.7.2)\cite{53}, concerning negative energy density of gravitational field (`Maxwell's Enigma' as Sattinger puts it) in a linear Lorentz invariant field theory of gravity are also refuted by Sattinger \cite{47}, who considered negative field energy density for SRMG in agreement with our result \cite{3}. In the discussion on the Dark Matter problem, Sattinger further noted: ``{\it The Maxwell-Heaviside equations of gravitation constitute a linear, relativistic correction to Newton's equations of motion; they interpolate between Newton's and Einstein's theories of gravitation, and are  therefore a natural mathematical model on which to build a dynamical theory of galactic structures}". 
Our unique, important and new findings reported and discussed here, again confirmed by our very recent work on ``Attractive Heaviside-Maxwellian (vector) Gravity from Quantum Field Theory" \cite{49} (where gravitational energy density for free fields is fixed positive by choice to address the objection of MTW (Sec. 7.2)\cite{53} without any inconsistency with the field equations (81-83) of SRMG), corroborate all the above suggested or derived linear vector gravitational equations in flat space-time which are seen to satisfy the correspondence principle (cp) in its correct sense: Newtonian Gravity 
$\Leftrightarrow$ NRMG $\Leftrightarrow$ SRMG. This means that the field equations of SRMG have room for both positive and negative energy solutions - a discussion of which is left out here. 
\subsection{Maxwellian Gravity From GR (GRMG)}
Different Maxwell-Lorentz-type equations for gravity obtained from GR by different researchers following different linearization procedures are not isomorphic \cite{50} as seen below: some contain non-linear terms and do not satisfy the cp from the perspective of our present new NRGM results substantiated by the cp-respecting SRMG results. Some samples of this type of General Relativistic Maxwellian Gravity (GRMG) are listed below for discussion.
\subsubsection{GRMG of Braginsky et al. and Forward (GRMG-BF):}
Braginsky et al. \cite{51}, following Forward \cite{52} and Misner, Thorne and Wheeler \cite{53}, reported the following Maxwell-type equations of GR in their parametrized-post-Newtonian (PPN) formalism as\footnote{Here we use the notation $\vec{E}_g$ for $\vec{g}$ and $\vec{H}_g$ for $\vec{H}$ in \cite{51}.}:
\begin{subequations}
\begin{align}
\mathbf{\nabla}\cdot\mathbf{E}_g =\,-\,4\pi G\rho_0\left[1+2\frac{\vec{v}^2}{c^2}+\frac{\Pi}{c^2}+\frac{3p}{\rho_0c^2}\right]   \nonumber \\
                   +\,\frac{3}{c^2}\frac{\partial^2\phi_g}{\partial t^2}     \\
\mathbf{\nabla}\times\mathbf{H}_g\,=\,-\,\frac{16\pi G}{c}(\rho_0\mathbf{v})\,+\,\frac{4}{c}\frac{\partial \mathbf{E}_g}{\partial t}  \\
\mathbf{\nabla}\cdot\mathbf{H}_g\,=\,0 \\
\mathbf{\nabla}\times\mathbf{E}_g\,=\,-\,\frac{1}{c}\frac{\partial \mathbf{H}_g}{\partial t}
\end{align}
\end{subequations}
\noindent 
where we have put the values of PPN parameters as appropriate for GR, $\rho_0$ is the density of rest mass in the local rest frame of the matter, $\vec{v}$ is the ordinary (co-ordinate velocity) velocity of the rest mass relative to the PPN frame, $\Pi$ is the specific internal energy (energy per unit rest mass) and $p$ is the radiation pressure and $\phi_g$ is the electric-type scalar potential. In terms of $(\phi_g)$ and magnetic-type gravitational vector potential $(\vec{A}_g)$, Braginsky et al. \cite{51} wrote (in our present notation)
\begin{eqnarray}
\vec{E}_g = -\vec{\nabla}\phi_g \,-\,\frac{1}{c}\frac{\partial\vec{A}_g}{\partial t},  \\
\vec{B}_g\,=\,\vec{\nabla}\times \vec{A}_g \\
\vec{\nabla}\cdot\vec{A}_g\,+\,\frac{3}{c}\frac{\partial \phi_g}{\partial t}\,=\,0 \qquad {(\mbox{For Lorenz-type Gauge})}
\end{eqnarray}
where the number $3$ in the Lorenz-type gauge above is the GR value for some PPN parameters used in \cite{51}. For the source and particle of velocities $|\vec{v}_0|\,<\, 10^5\,\mbox{cm/sec} << \,c$ , Braginsky et al. \cite{51} approximated the gravitational force (with a typographical error in eqn. (3.10)\footnote{Viz.: $\frac{\vec{F}}{m}=\left[1+\frac{1}{2}(2\gamma+1)\right]\frac{\vec{v}^2}{c^2}\vec{E}_g+\frac{1}{c}\left(\vec{v}\times\vec{H}_g\right)$.}, p.2054, \cite{51}, corrected here) on a unit mass,   
\begin{equation}
\frac{\vec{F}}{m_0}= \left[1+\frac{1}{2}(2\gamma+1)\frac{\vec{v}_0^2}{c^2}\right]\vec{E}_g+\frac{1}{c}\left(\vec{v_0}\times\vec{H}_g\right),
\end{equation}
\noindent 
where the PPN parameter $\gamma \simeq 1$ in GR. In empty space ($\rho_0 = 0$) with no radiation pressure ($p = 0$), if we consider Coulomb-Newton Gauge ($\vec{\nabla}\cdot\vec{A}_g = 0$), the field equations (92a-92d) reduce to the following equations
\begin{subequations}
\begin{align}
\mathbf{\nabla}\cdot\mathbf{E}_g =\,0,   \\
\mathbf{\nabla}\times\mathbf{H}_g\,=\,+\,\frac{4}{c}\frac{\partial \mathbf{E}_g}{\partial t},  \\
\mathbf{\nabla}\cdot\mathbf{H}_g\,=\,0, \\
\mathbf{\nabla}\times\mathbf{E}_g\,=\,-\,\frac{1}{c}\frac{\partial \mathbf{H}_g}{\partial t}.
\end{align}
\end{subequations}
Now taking the curl of (97d) and utilizing equations (97a) and  (97b), we get the wave equation for the field $\vec{E}_g$ in empty space as
\begin{equation}
\vec{\nabla}^2\vec{E}_g\,-\,\frac{4}{c^2}\frac{\partial ^2\vec{E}_g}{\partial t^2} = \vec{\nabla}^2\vec{E}_g\,-\,\frac{1}{c_g^2}\frac{\partial ^2\vec{E}_g}{\partial t^2} =\vec{0},
\end{equation}
where $c_g\,=\,c/2$. Similarly, the wave equation for the field $\vec{H}_g$ can be obtained by
taking the curl of equation (97b) and utilizing equations (97c) and (97d): 
\begin{equation}
\vec{\nabla}^2\vec{H}_g\,-\,\frac{4}{c^2}\frac{\partial ^2\vec{H}_g}{\partial t^2} = \vec{\nabla}^2\vec{H}_g\,-\,\frac{1}{c_g^2}\frac{\partial ^2\vec{H}_g}{\partial t^2} =\vec{0},
\end{equation}
where again we find $c_g\,=\,c/2$. This result is against the special relativistic (as well as
the gauge field theoretic) expectation that the speed of gravitational waves (if they
exist) should be equal to the speed of light in any Lorentz-covariant field theory of
gravity and has escaped the attention of the authors \cite{51}. It is to be noted that the odd factor of $4$ is responsible for this strange result. Thus GRMG-BF formulation is defective not only for yielding $c_g = c/2$ for gravitational waves in vacuum but also for the Goravito-Lorentz force law not satisfying satisfying the correspondence principle as judged from the 
the non-relativistic Gravito-Lorentz force of NRMG or of the SRMG. Further equation of continuity does not follow from GRMG-BF field equations because of the existence of some non-linear terms in equation (92a). 
\subsubsection{GRMG of Harris (GRMG-H):}
Instead of using the PPN formalism of GRMG-BF, Harris \cite{54} derived a set of gravitational equations for slowly moving particles in weak gravitational fields starting from the equations of GR. The resulting equations have some resemblance to those in electromagnetism: 
\begin{subequations}
\begin{align}
\mathbf{\nabla}\cdot\mathbf{E}_g =\,-\,4\pi G\rho_0,    \\
\mathbf{\nabla}\times\mathbf{H}_g\,=\,-\,\frac{16\pi G}{c}(\rho_0\mathbf{v})\,+\,\frac{4}{c}\frac{\partial \mathbf{E}_g}{\partial t},  \\
\mathbf{\nabla}\cdot\mathbf{H}_g\,=\,0, \\
\mathbf{\nabla}\times\mathbf{E}_g\,=\,-\,\frac{1}{2c}\frac{\partial \mathbf{H}_g}{\partial t}.
\end{align}
\end{subequations}
The Gravito-Lorentz force equation of Harris is of the following form
\begin{equation}
m_0\frac{d\vec{v}}{dt}\,=\,m_0\left[\vec{E}_g\,+\frac{1}{c}(\vec{v}\times\vec{H}_g)\right]\,+\,m_0\vec{v}\frac{1}{2c^2}\frac{\partial \phi_g}{\partial t}.
\end{equation}
The field $\vec{E}_g$ is related to the scalar potential ($\phi_g$) and vector potential ($\vec{A}_g$) as 
\begin{equation}
2\mathbf{E}_g\,=\,-\,\mathbf{\nabla}\phi_g\,-\,\frac{1}{c}\frac{\partial \mathbf{A}_g}{\partial t}.
\end{equation} 
In empty space (where $\rho_0 = 0$), the field equations (100a-100d) give us the following 
wave equations for the fields ($\mathbf{E}_g, \mathbf{B}_g$): 
\begin{subequations}
\begin{align}
\mathbf{\nabla}^2\mathbf{E}_g\,-\,\frac{2}{c^2}\frac{\partial ^2\mathbf{E}_g}{\partial t^2} = \mathbf{\nabla}^2\mathbf{E}_g\,-\,\frac{1}{c_g^2}\frac{\partial ^2\mathbf{E}_g}{\partial t^2} =\mathbf{0},  \\
\mathbf{\nabla}^2\mathbf{H}_g\,-\,\frac{2}{c^2}\frac{\partial ^2\mathbf{H}_g}{\partial t^2} = \mathbf{\nabla}^2\mathbf{H}_g\,-\,\frac{1}{c_g^2}\frac{\partial ^2\mathbf{H}_g}{\partial t^2} =\mathbf{0},
\end{align}
\end{subequations}
\noindent
where $c_g\,=\,c/\sqrt{2}$. So GRMG-H is also defective like GRMG-BF. 
\subsubsection{GRMG of Ohanian and Ruffini (GRMG-OR)}
In the Non-relativistic limit and Newtonian Gravity corespondence of GR, Ohanian and Ruffini \cite{20} (Sec. 3.4 of \cite{20}) obtained the following equations from GR: 
\begin{equation}
\frac{d\mathbf{v}}{dt} = \mathbf{g}\,+\,\mathbf{v}\times \mathbf{b} 
\end{equation} 
\begin{subequations}
\begin{align}
\mathbf{\nabla}\cdot\mathbf{g}\,=\,-\,4\pi G \rho_0  \\
\mathbf{\nabla}\times\mathbf{g}\,=\,-\,\frac{1}{2}\frac{\partial \mathbf{b}}{\partial t} \\
\mathbf{\nabla}\cdot\mathbf{b}\,=\,0  \\ 
\mathbf{\nabla}\times\mathbf{b}\,=\,-\,\frac{16\pi G}{c^2} \mathbf{j}_0 
\end{align}
\end{subequations} 
where $\rho_0$ is the (rest) mass density, $\mathbf{j}_0$ is the momentum density. The equation (105d) 
(representing the gravito-Amp\`{e}re law) is valid for time independent field \cite{20}, i.e.,
$\partial \mathbf{g}/\partial t = \mathbf{0}$, which in view of equation (105a) is equivalent 
to $\partial \rho_0/\partial t = 0$. Since the divergence of curl of any vector is identically zero, 
the divergence of equation (105d) gives us $\mathbf{\nabla}\cdot \mathbf{j}_0 = 0$. Thus equation 
(105d) has the same limitation as that of the Amp\`{e}re's law of electromagnetism. Therefore, it needs a correction like Maxwell's correction to the Amp`{e}re's law. While the conditions $\mathbf{\nabla}\cdot \mathbf{j}_0 = 0$ and $\partial \rho_0/\partial t = 0$ are valid for steady-state problems, the general situation, where $\mathbf{\nabla}\cdot \mathbf{j}_0 \neq 0$ and $\partial \rho_0/\partial t \neq 0$, is given by the continuity equation for mass and mass current or momentum density: 
\begin{equation}
\mathbf{\nabla}\cdot \mathbf{j}_0 +\frac{\partial \rho_0}{\partial t}\,= 0.
\end{equation}
The equations (105a) and (105d) will be consistent with the continuity equation (106), if we make the following Maxwell-like correction to the gravito-Amp\`{e}re law (105d): 
\begin{equation}
\mathbf{\nabla}\times\mathbf{b} = - \frac{16\pi G}{c^2} \mathbf{j}_0 +\frac{4}{c^2}\frac{\partial \mathbf{g} }{\partial t}. 
\end{equation}
Further, without the above correction to the equation (105d), there can not be gravitational waves. Now, the corrected self-consistent field equations (105a-105c, 107) yield transverse gravitational waves; the wave equations for the $\mathbf{g}$ and $\mathbf{b}$ fields of GRMG-OR, in vacuum, take the following forms:
\begin{subequations}
\begin{align}
\mathbf{\nabla}^2\mathbf{g}\,-\,\frac{2}{c^2}\frac{\partial ^2\mathbf{g}}{\partial t^2} = \mathbf{\nabla}^2\mathbf{g}\,-\,\frac{1}{c_g^2}\frac{\partial ^2\mathbf{g}}{\partial t^2} =\mathbf{0},  \\
\mathbf{\nabla}^2\mathbf{b}\,-\,\frac{2}{c^2}\frac{\partial ^2\mathbf{b}}{\partial t^2} = \mathbf{\nabla}^2\mathbf{b}\,-\,\frac{1}{c_g^2}\frac{\partial ^2\mathbf{b}}{\partial t^2} =\mathbf{0},
\end{align}
\end{subequations}
\noindent
where $c_g\,=\,c/\sqrt{2}$. This means that the transverse gravitational waves originating from slowly varying fields and weak sources travel through vacuum not at the light speed but at a reduced speed $c_g\,=\,c/\sqrt{2}$. Thus GRMG-OR will have exact correspondence with NRMG provided $c_g\,=\,c/\sqrt{2}$. However, GRMG-OR will not correspond to the slow motion approximation of SRMG where $c_g = c$ in vacuum even at relativistic motion of the fields and sources. Now, if we define a new filed
for GRMG-OR as 
\begin{equation}
\mathbf{\tilde{b}} = \frac{\mathbf{b}}{2},
\end{equation}   
then the gravito-Lorentz force law of GRMG-OR will take the form:
\begin{equation}
\frac{d\mathbf{v}}{dt} = \mathbf{g}\,+\,2\mathbf{v}\times \mathbf{\tilde{b}} 
\end{equation}
and the field equations (105a-105c, 107) will take the form 
\begin{subequations}
\begin{align}
\mathbf{\nabla}\cdot\mathbf{g}\,=\,-\,4\pi G \rho_0  \\
\mathbf{\nabla}\times\mathbf{g}\,=\,-\,\frac{\partial \mathbf{\tilde{b}}}{\partial t} \\
\mathbf{\nabla}\cdot\mathbf{\tilde{b}}\,=\,0  \\ 
\mathbf{\nabla}\times\mathbf{\tilde{b}} = - \frac{8\pi G}{c^2} \mathbf{j}_0 +\frac{2}{c^2}\frac{\partial \mathbf{g} }{\partial t}
\end{align} 
\end{subequations}
Again these equations yield wave equations for the $\mathbf{g}$ and $\mathbf{\tilde{b}}$ fields for which $c_g\,=\,c/\sqrt{2}$. Further, if we define $\mathbf{B}_g = \mathbf{b}/4$, the redefined field equations (not written here) yield wave equations for the $\mathbf{g}$ and $\mathbf{B}_g$ fields for which $c_g\,=\,c/\sqrt{2}$ in vacuum, while the gravito-Lorentz force law of GRMG-OR takes the following form 
\begin{equation}
\frac{d\mathbf{v}}{dt} = \mathbf{g}\,+\,4\mathbf{v}\times \mathbf{B}_g. 
\end{equation}
The defects of GRMG-OR are apparent from the perspectives of cp violation and the suspected value of $c_g$ in Einstein's linearized equations. 
\subsubsection{GRMG of Pascual-S\`{a}nchez, Moore, (GRMG-PS-M):}
Following Huei \cite{55}, Wald \cite{56} and Ohanian and Ruffini \cite{57}, Pascual-S\`{a}nchez \cite{58}, by using some approximation of GR, obtained the following set of Lorentz-Maxwell-like gravitomagnetic equations that match with Moore's \cite{59} equations from GR in the the weak field and slow motion approximation:  
\begin{equation}
m_0\frac{d\vec{v}}{dt}\,=\,m_0\left(\vec{E}_g\,+\,4\vec{v}\times\vec{B}_g\right).
\end{equation}
\begin{subequations}
\begin{align}
\mathbf{\nabla}\cdot\mathbf{E}_g =\,-\,4\pi G\rho_0,    \\
\mathbf{\nabla}\times\mathbf{B}_g\,=\,-\,\frac{4\pi G}{c^2}(\rho_0\mathbf{v})\,+\,\frac{1}{c^2}\frac{\partial \mathbf{E}_g}{\partial t},  \\
\mathbf{\nabla}\cdot\mathbf{B}_g\,=\,0, \\
\mathbf{\nabla}\times\mathbf{E}_g\,=\,-\,\frac{\partial \mathbf{B}_g}{\partial t}.
\end{align}
\end{subequations}
Although the field equations (114a - 114d) match with those of GRMG-UG \cite{1}, SRMG where $c_g = c$ exactly and with NRMG equations conditionally when $c_g = c$ as shwon here, the Gravito-Lorentz force, containing a factor of $4$ in the gravitomagnetic interaction, does not satisfy the cp in the sense discussed here. It is to be noted that Ciubotariu \cite{60} considered Peng's \cite{61} version of GRMG equations\footnote{which, may be of GRMG-PS-M type as the author could not get the ref. \cite{61} and Ciubotariu.} in the prediction of absorption of gravitational waves, while Minter et al. \cite{62} considered GRMG-PS-M version in their investigation on the question of the existence of mirrors for gravitational waves. 
\subsubsection{Maxwellian Gravity of Mashhoon (MG-Mashhoon):}
Mashhoon \cite{63,64}, in his general linear solution of the gravitational field equations of Einstein, obtained following gravitational analogues of Maxwell's equations: 
\begin{subequations}
\begin{align}
\mathbf{\nabla}\cdot\mathbf{E}_g =\,4\pi G\rho_0,    \\
\mathbf{\nabla}\times\left(\frac{1}{2}\mathbf{B}_g\right)\,=\,\frac{4\pi G}{c}(\rho_0\mathbf{v})\,+\,\frac{1}{c}\frac{\partial \mathbf{E}_g}{\partial t},  \\
\mathbf{\nabla}\cdot\left(\frac{1}{2}\mathbf{B}_g\right)\,=\,0, \\
\mathbf{\nabla}\times\mathbf{E}_g\,=\,-\,\frac{1}{c}\frac{\partial }{\partial t}\left(\frac{1}{2}\mathbf{B}_g\right).
\end{align}
\end{subequations}
By defining $\vec{E}_g$ and $\vec{B}_g$ fields in terms of scalar and vector potentials $(\phi_g, \vec{A}_g)$ as
\begin{equation}
\vec{E}_g = - \vec{\nabla}\phi_g - \frac{1}{c}\frac{\partial}{\partial t}\left(\frac{1}{2}\vec{A}_g\right), \quad \vec{B}_g\,=\,\vec{\nabla}\times\vec{A}_g, 
\end{equation}
he wrote the Lagrangian, L, for the motion of a test particle of rest mass $m_0$ (to linear order in $\phi_g$ and $\vec{A}_g$) as
\begin{equation}
\begin{split}
L &= -m_0c^2\left(1-\frac{v^2}{c^2}\right)^{\frac{1}{2}}  \\
                                                     & \quad +m_0 \gamma\left(1+\frac{v^2}{c^2}\right)\phi_g - \frac{2m_0}{c}\gamma \vec{v} \cdot \vec{A}_g,
\end{split}
\end{equation}
where $\gamma = (1 - v^2/c^2)^{-1/2}$ is the Lorentz factor. The equation of motion, 
$d\vec{p}/dt =\vec{F}$, where $\vec{p}=\gamma m\vec{v}$ is the kinetic momentum, is expressed as 
\begin{equation}
\frac{d\vec{p}}{dt} = - m_0\vec{E}_g -2m_0\frac{\vec{v}}{c}\times \vec{B}_g,
\end{equation}
if $\partial \vec{A}_g/\partial t = \vec{0}$ and $\vec{F}$ is expressed to lowest order in 
$v/c$, $\phi_g$ and $\vec{A}_g$.\\
In empty space, the field equations (115a-115d) give us wave equations for $\vec{E}_g$ and $\vec{B}_g$ fields with $c_g = c$. But because of the factor of $2$ appearing in the relativistic Gravito-Lorentz force law (118), it does not correspond to the cp respecting special relativistic Gravito-Lorentz force law of SRMG expressed in equations (77, 83).
\begin{equation}
\frac{d\vec{p}}{dt} = - m_0\vec{E}_g - m_0\frac{\vec{v}}{c}\times \vec{B}_g,
\end{equation}
written here in Mashhoon's convention for the gravitostatics Gauss's law (115a) and Gravito-Lorentz force Law (118). Further, in the non-relativistic case also, the equation (118) does not match with our non-relativistic result here, if $c_g = c$. \\
In our above discussions, we found that the speed of gravitational waves in different linearized versions of Einstein's equations is not unique; it depends on the thought of a scientist concerning the mangagement of the spurious factor of 4 by splitting it $4=2\times2$ and moving a factor of $2$ to some other place of the equations for consistency\footnote{Such a scheme my be made general by thinking $4 = 2x\times \frac{2}{x}$ with $x$ being any non-zero real number to get other sorts of spurious results in a self-consistent way.}. Eddington \cite{65} in his textbook ``{\it The Mathematical Theory of Relativity}, which Einstein suggested was ``the finest presentation of the subject in any language", rightly found and said that weak-field solutions of the wave equation obtained from Einstein's field equations were just coordinate changes which we can ``propagate" with the {\it speed of thought}. Apart from this defect of yielding a non-unique value of $c_g$, the linearized theory of gravitational waves has its limits because the linear approximation is not valid for sources where gravitational self-energy is not negligible \cite{66}, as in the case of merging of highly compact objects like the Neutron stars and the `so called' Black Holes. It is important to note that the current experimental data on observation of gravitational waves \cite{6,7,8,9} and grvitomagnetic phenomena are being (or may be) explained by using any one of these cp-defying linearized versions of GR.
Further, these generally perceived general relativistic phenomena are being interpreted as having no counterpart in the Newtonian world, which we found not to be true and satisfactory. The author wishes to stress that just by connecting the Gauss Law of gravitostatics and equation of continuity in a consistent way, one gets gravitational analogue of Amp\`{e}re-Maxwell law, which is being hailed as one of the most important predictions of GR (where the mathematical trees seem to obscure the physical forest). Any talk of gravitomagnetism has long been the prerogative of general relativists. With our present report, even undergraduates, untrained in the mathematical gymnastics of general relativity, can now talk and think of the generation, transmission and detection of gravitational waves \cite{67,68,69}\footnote{Which, in the framework of gravito-electromagnetic theory, cannot be just any other wave, but the waves whose nature is dictated by equations of gravitodynamics: Gravito-Maxwell Equations. By contrast, the general description of gravitational waves in GR is different \cite{66}. However, in spite of the interesting work by Mead \cite{67}, attempting to provide a non-general relativistic explanation of the generation and detection of gravitational waves, which Isi et al. \cite{68} referred to, a proper description of gravitational waves within a purely gravitoelectromagnetic formalism is still far from being established.} akin to electromagnetic waves as matter of scientific curiosity and investigate the role of gravitoelectromagnetism in different fields of study, ranging from classical physics to quantum physics \cite{1,3} and to cosmology \cite{36,42} even quantum cosmology to test the validity (or the domain of validity) of the proposed gravitoelectromagnetic theory from experimental point of view. \\
Basically, we offer a new pack of beautiful cards (or a beautifully simple, self-consistent toy model theory of gravity) to play with, if one likes. However, it remains to be seen how the recently observed phenomena concerning gravitomagnetism and gravitational waves may be interpreted with our new findings with $c_g = c$, since these phenomena could well be explained by NRMG just by adjusting the undetermined parameter to $c_g = c/2\,\, {\mbox {or}}\,\, c/\sqrt{2}$ or by cp-violating SRMG without invoking the curved space-time concept of Einstein and keeping other experimental parameters intact or by SRMG ($c_g = c$ with normal gravito-Lorentz force) and varying other parameters of the experiment in a self consistent way, or re-analyzing the sources of possible theoretical and experimental errors in the interpretation of the experimental data. The author aims to address these questions in future, if his odd situation permits and wishes the young minds do this, if they can. 
\subsection{Spin-1 Vector Gravity vs Spin-2 Tensor Gravity}
Many quantum field theorists, like Gupta \cite{70}, Feynman \cite{71}, Zee \cite{72} and Gasperini \cite{73}, to name a few, have rejected spin-1 vector theory of gravity on the ground that if gravitation is described by a vector field theory like Maxwell's electromagnetic theory as discussed here, then vector-like interactions will produce repulsive static interactions between sources of the same sign, while - according to Newton's gravitational theory - the static gravitational interaction between masses of the same sign is attractive.  However, in one of our recent work \cite{49} on quantum field theoretical rediscovery of Heaviside-Maxwellian gravity (in flat space-time) (same as SRMG), we have shown that this not true because of the existence of a fundamental difference in the sign before the source term in the in-homogeneous field equations of SRMG and relativistic Maxwell's electromagnetism (RMEM) as seen below in SI units: 
\begin{equation}
\Box A_{g}^\mu = - \mu_{0g}j_0^\mu, \quad \quad \Box A^\mu = \mu_{0}j_e ^\mu  
\end{equation}
where $\Box = \partial ^\mu \partial _\mu = \frac{1}{c^2}\frac{\partial ^2}{\partial t^2} - \mathbf{\nabla}^2$, $A_{g}^\mu$ and $j_0^\mu$ respectively represents the 4-potential and 4-mass current density of SRMG while $A^\mu$ and $j_e^\mu$ respectively represents the 4-potential and 4-charge current density of RMEM. This fact that a vector gravitational theory, as proposed first by Heaviside and later rediscovered by many others following different appoaches, implies attractive interaction between static masses was transparently clear to Sciama \cite{36}. In spin-1 SRMG, contrary to the electromagnetic cases, like masses (or gravitational charges) should attract and unlike masses (if they exist) should repel each other under static conditions, while like (parallel) mass currents repel and un-like mass currents attract each other \cite{3}. Similarly, there should be attraction between like gravitomagnetic poles and repulsion between un-like gravitomagnetic poles: opposite to the case in electromagnetism where like magnetic poles repel and un-like magnetic poles attract each other \cite{3}. Following Dirac's scheme of predicting the spin magnetic moment of an electron, in our previous work \cite{3}, we have shown that the gravitomagnetic moment of a Dirac (spin $1/2$) fermion is exactly equal to its spin angular momentum: $\mu_{sg} = \frac{\hbar}{2}$, which can just be inferred from the magnetic moment of a Dirac (spin $1/2$) fermion, $\mu_{se} = \frac{q\hbar}{2m_0}$, by replacing the electric charge $q$ with the  gravitational charge $m_0$ of the Dirac fermion as per SRMG. However, in GR, $\mu_{sg}$ value is not unique as found by different authors and noted in \cite{3}.    \\
Regarding the idea of spin-2 graviton, Wald \cite{56} noted that the linearized Einstein's equations in vacuum are precisely the equations written down by Fierz and Pauli \cite{74}, in 1939, to describe a massless spin-2 field propagating in flat space-time. {\it Thus, in the linear approximation, general relativity reduces to the theory of a massless spin-2 field which undergoes a non-linear self- interaction. It should be noted, however, that the notion of the mass and spin of a field require the presence of a flat back ground metric $\eta_{ab}$ which one has in the linear approximation but not in the full theory, so the statement that, in general relativity, gravity is treated as a mass-less spin-2 field is not one that can be given precise meaning outside the context of the linear approximation} \cite{56}. Even in the context linear approximation, the original idea of spin-2 graviton gets obscured due to the several faces of Gravito-Maxwell equations seen here. This may be seen as another limitation of GR for not making a unique and unambiguous prediction on the spin of graviton.   \\  
\subsection{Little Known Heaviside's Work on Gravity}
Heaviside's work on gravity, which McDonald \cite{4} called {\it a low velocity, weak-field approximation to general relativity}, is little known and has not received as much attention as it deserves. This is because, in many leading papers and books exploring gravitomagnetic phenomena and gravitational waves, one finds rare or no mention of Heaviside's name, although Heaviside \cite{5} predicted gravitomagnetic effects and considered the necessity of possible existence of gravitational waves (for which there must be some wave equations first written down by him) almost 20 years before Einstein's prediction of gravitational waves \cite{75,76}. Further, it is more surprising not to find Heaviside's name in the {\it Scientific Background on the Nobel Prize in Physics 2017} \cite{77} when the 2017 Physics Nobel Prize was declared to be awarded to Rainer Weiss, Barry C. Barish and Kip S. Thorne  {\it for ({\it their}) decisive contributions to the LIGO detector and the observation of gravitational waves}. Brillouin \cite{38}, in his final remark on Carstoiu's \cite{37} suggestions for gravity waves aptly stated, ``{\it It is very strange that such an important paper had been practically ignored for so many years, but the reader may remember that Heaviside was the forgotten genius of physics, abandoned by everybody except a few faithful friends.}"  
\section{Conclusions} 
Following Schwinger's non-relativistic formalism of classical electrodynamics, here we derived the fundamental equations of Non-Relativistic Maxwellian Gravity (NRMG), which matches with Heaviside's Gravity of 1893 and offers a plausible mechanism for resolving the problem of action-at-a-distance in Newtonian gravity, within Galileo-Newtonian domain of physics by demanding the existence of gravitational waves propagating in vacuum at a non-zero finite speed $c_g$, whose value has to be determined from experiments on measurable quantities involving $c_g$ or from some more advanced theory. Then, following an independent special relativistic approach, we re-discovered NRMG in its relativistic version and named it Special Relativistic Maxwellian Gravity (SRMG), where $c_g = c$ comes out naturally 
and the equality of gravitational mass $m_g$ with Lorentz-invariant rest mass $m_0$ is clearly demonstrated (resolving Eddington's ``mass ambiguity") in the re-examination of an old thought experiment aimed at finding the natural conditions of equilibrium of a two particle system having requisite masses and electric charges in two inertial frames in relative motion such that the equilibrium remains frame-independent. Most importantly the equality $m_g = m_0$ emerges as a consequence of the Lorentz-invariance of physical laws and the Law of Universality of Free Fall emerges as a consequence of $m_g = m_0$, not an initial assumption in SRMG. Both NRMG and SRMG obey correspondence principle (cp) in its true sense: Newtonian Gravity $\Leftrightarrow$ NRMG $\Leftrightarrow$ SRMG. These flat space-time versions of Maxwellian gravity matches with those considered by several authors either for explaining some GR tests or for their derivation of SRMG following different approaches. By the way, we also considered a non-linear form of SRMG (SRMG-N) in flat spacetime, where the non-linearity arises due to the initial axiom of $m_g = m_0/\sqrt{1 - v^2/c^2}$. The non-relativistic linearized version of SRMG-N does not correspond to NRMG when $c_g = c$ or the non-relativistic version of SRMG. Thus, SRMG-N seems to defy the cp. Further, we noted several versions of General Relativistic Maxwellian Gravity (GRMG), including Ummarino and Gallerati's version, which seem to defy the cp: Newtonian Gravity $\Leftrightarrow$ NRMG $\Leftrightarrow$ SRMG $ \cancel{\Leftrightarrow}$ GRMG; although they are being or may be (rightly or wrongly) employed to explain the experimental data on gravitational waves and the whole other class of gravitomagnetic effects predicted by GR. While SRMG unambiguously fixes the exact value of $c_g = c$ and spin of graviton $s_g = 1$ uniquely, their values in GR are ambiguous and non-unique. However, the author leaves it for the consideration of the readers to decide which version of gravitoelctromagnetism or Maxwellian Gravity is to be taken into consideration not only in the interpretation, theoretical as well as experimental error analysis of recent experimental data on the detection of gravitomagnetic field generated by mass-energy currents and the very recent detection of gravitational waves but also in the search for the interplay of gravitational fields with other fields/states of matter in nature. The author wishes to make an important remark that none of the authors who come up with a factor of 4 or 2 in their Lorentz-Maxwell-like solutions of GR have made an error in their calculations; the spurious factor of $4$ or $2$ (surprisingly $1$ in GRMG-UG formulation), ``really does" follow from GR or rather from its basic building blocks: Einstein's initial axioms of (i) $m_g = E/c^2$ and (ii) space-time curvature, taken as inputs to the whole mathematical structure of GR. In our discussion of non-linear SRMG (SRMG-N) in flat space-time, we have shown that a factor of $2$ originates form the adoption of $m_g = E/c^2$ in flat (Minkowski) spacetime, hence the origin of another factor of ``$2$" or ``$1$" in linearized GR may be attributed to the adoption of the notion of space-time curvature and on how one chooses to define the gravitomagnetic field in terms of the potentials (that is, the perturbations in the metric) in different linearization schemes. Moreover, our findings in no way affect the main conclusions of Ummarino and Gallerati's paper on ``Superconductor in a weak static gravitational field".
\begin{acknowledgements}
The author is extremely grateful to the anonymous referee and author's PhD supervisor Prof. Gautam Mukhopadhyay, Department of Physics, IIT Bombay, Mumbai, India, for their many valuable suggestions for the improvement of the manuscript. The author thanks his wife, Pramila, for her constant support and encouragement in this endeavor. His sincere thanks are also due to Prof. Niranjan Barik, Physics Department, Utkal University, Vanivihar, Bhubaneswar, Odisha, India, for his constant motivation to the author to pursue things from new perspectives in consonance with the scientific philosophy of Einstein and Infeld (in: The Evolution of Physics):``{\it To raise new questions, new possibilities, to regard old problems from a new angle, requires creative imagination and marks real advance in science}." 
\end{acknowledgements}



\end{document}